\documentclass[12pt,preprint]{aastex}
\usepackage{psfig}

\def\gs{\mathrel{\raise0.35ex\hbox{$\scriptstyle >$}\kern-0.6em 
\lower0.40ex\hbox{{$\scriptstyle \sim$}}}}
\def\ls{\mathrel{\raise0.35ex\hbox{$\scriptstyle <$}\kern-0.6em 
\lower0.40ex\hbox{{$\scriptstyle \sim$}}}}
\def\et{\hbox{et al.}$\,$}
\def\OII{\hbox{[O II]}}
\def\Hd{\hbox{H$\delta$}}
\def\EWOII{\hbox{EW([O II])}}
\def\EWHd{\hbox{EW(H$\delta$)}}

\def\kms{\rm{\hbox{km s$^{-1}$}}}
\slugcomment{Astrophysical Journal -- submitted: 2004 March 5; accepted: 
2004 August 3 }

\lefthead{Dressler et al.}
\righthead{Studying Star Formation Histories...}

\received{2004 March 5}
\begin{document}

\title{Studying The Star Formation Histories of Galaxies in Clusters from 
Composite Spectra} 

\author{Alan Dressler \& Augustus Oemler, Jr.}
\affil{Carnegie Observatories, 813 Santa Barbara St., Pasadena, California 
91101-1292}
\email{dressler@ociw.edu, oemler@ociw.edu}

\author{Bianca M.\ Poggianti} 
\affil{Osservatorio Astronomico di Padova, vicolo 
dell'Osservatorio 5, 35122 Padova, Italy}
\email{poggianti@pd.astro.it}

\author{Ian Smail}
\affil{Department of Physics, University of Durham, South Rd, 
Durham DH1 3LE, UK}
\email{Ian.Smail@durham.ac.uk}

\author{Scott Trager}
\affil{ Kapteyn Institute, Rijksuniversiteit Groningen, 
Postbus 800, NL-9700 AV Groningen, Netherlands}
\email{sctrager@astro.rug.nl}

\author{Stephen A. Shectman}
\affil{Carnegie Observatories, 813 Santa Barbara St., Pasadena, California 
91101-1292}
\email{shec@ociw.edu}

\author{Warrick J.\ Couch}
\affil{School of Physics, University of New South Wales, Sydney 2052, 
Australia, w.couch@unsw.edu.au}

\author{Richard S.\ Ellis}
\affil{California Institute of Technology, Pasadena, California, rse@astro.caltech.edu}

\begin{abstract}

We have formed ``composite spectra'' by combining the integrated--light
spectra of individual galaxies in 8 intermediate--redshift and 12
low--redshift clusters of galaxies.  Because these composite spectra
have much higher signal--to--noise ratios than individual galaxy spectra,
they are particularly useful in quantifying general trends in star
formation for galaxy populations in distant clusters, $z > 0.3$.  
By measuring diagnostic features that represent stellar populations of
very different ages, a grand-composite spectrum can reflect the 
fractions of those populations as accurately as if excellent
spectral measurements were available for each galaxy. Such
composite spectra can also be useful in the study of finer spectral
signatures, for example, spectral indices that break the
age--metallicity degeneracy, and the shape of the \Hd\ absorption line
as an indicator of the age and duration of an epoch of starbursting
galaxies in a cluster.

Measuring the equivalent widths of spectral features in composite
spectra is especially well--suited for comparing cosmic variance of
star formation in clusters at a given redshift, or comparing clusters
over a range of redshifts.  When we do this we find that [O II]
emission and especially Balmer absorption is strong in each of our
intermediate--redshift clusters, and completely separable from a sample
of 12 {\it present--epoch} clusters, where these features are weak.
Cluster--to--cluster variations at a given epoch seem to be smaller than
the strong trend in redshift, which suggests that cosmic evolution is
the major factor in the star formation histories of cluster galaxies.
Specifically, we show by comparing to the \Hd\ strengths of
present--epoch populations of continuously star--forming galaxies that
the higher--redshift samples must contain a much higher fraction of
starburst galaxies than are found today in any environment.

\end{abstract}

\keywords{galaxies: clusters: general --- galaxies: evolution}

\section{Introduction}

Butcher and Oemler's (1978a,b) report of strong color evolution of
cluster galaxies since $z = 0.5$ was unexpected within the context of
ideas about galaxy evolution that were prevalent at the time.  Studies
of the stellar populations of nearby galaxies had fostered the notion
that elliptical and S0 galaxies are basically old stellar systems; the
population of present--epoch clusters are so dominated by these types
that it seemed surprising that rich clusters would have been sites of
significant star formation only $\sim$4 Gyr ago.  Nevertheless, the
blue colors Butcher \& Oemler observed for one--third to one--half the
galaxies in the original two clusters of their study suggested just
that.  At that time, before hierarchical clustering became the popular
paradigm for galaxy evolution, the idea that these early type galaxies
might be assembled so recently --- scarcely one--third of a Hubble time
ago --- was a radical proposition.

In the following two decades, evidence for the Butcher--Oemler effect
accumulated (see, e.g., Margoniner \& de Carvalho 2000, Margoniner 
\et\ 2001) and the focus began to shift to spectroscopic studies
which could connect the Butcher--Oemler effect with the individual and
collective evolution of star forming galaxies in the cluster
environment.  Dressler \& Gunn (1982, 1983) were the first to confir
that the blue populations of the Butcher--Oemler clusters were the
result of star formation, and also to point out that the presence of
strong \Hd\ absorption indicated a greater prevalence of starburst
galaxies than are found today (Couch \& Sharples 1987; Giraud 1990;
Kelson \et\ 1997, 2000; Tran \et~2003).  Our present collaboration --- the
Morphs --- has obtained a large spectroscopic sample and presented
evidence for a strong spectral evolution of galaxies in rich clusters
since $z \sim 0.5$.  Our technique has been to use the spectral
features of individual galaxies to assign them different spectral
classes (Dressler \et\ 1999, D99; Poggianti \et\ 1999, P99), and compare
clusters using their fractions of the different spectral classes.  For
our ten--cluster intermediate--redshift sample, P99 used this approach to
conclude that, within counting statistics, all exhibit the same
behavior of enhanced ongoing star formation compared to present--epoch
clusters, and a much pronounced component of starburst
galaxies. However, other studies have reported on clusters that appear
to have much less ongoing or recent star formation, in particular,
much less in the form of starbursts (Balogh \et\ 1999).  The attempt to
resolve whether this is a genuine difference of cluster populations
has made clear that the barely adequate signal--to--noise ratio (S/N) of
the spectra in these studies, and the lack of agreement on a common
system of spectral classes, hampers efforts to arrive at a consistent
picture.

We present here a new technique of comparing clusters or any galaxy
populations that facilitates such comparisons, based on coadding all
available spectra in each cluster. The idea is to obtain a composite of
the starlight over the cluster, by adding spectra of {\it a
representative population of galaxies in the cluster}. For a sample of
clusters, if the available spectra sample the galaxy population to a
similar magnitude limit and magnitude distribution, and are unbiased
with respect to color or type, co--adding a sufficient number of spectra
will produce for each cluster a high S/N spectrum of the composite
stellar population.  In these composite spectra the absorption and
emission features that are used to assess star formation history are
very well defined, so the comparison between clusters, and between
studies, should be much easier to make, and more reliable.

For this study we have chosen to measure \OII\ emission and \Hd\
absorption in the composite spectra.  By this choice we are
essentially asking: (1) How much of the light in the cluster is
``old'' --- more than a few Gyr? (2) How much has an age 500 Myr to 1
Gyr --- the lifetimes of A--stars? (3) How much is from active regions
of star formation with characteristic ages of $10^{7}$ years? 
Because these three populations are temporally distinct, they provide
a sort of orthonormal basis set for dissecting the population
of a single galaxy or an entire cluster, and, since composite spectra 
have a high S/N ratio compared to the typical individual spectra, the 
results of such measurements are robust and unambiguous.  They 
furthermore offer the potential of measuring more subtle features 
that are important diagnostics of stellar population evolution.

In P99 we analyzed spectra of individual cluster galaxies with strong
\OII\ and \Hd\ and showed, by comparing to the models of continuously
star--forming galaxies, that in numerous cases the strength of these
features were the clear signature a starburst, or post--starburst star
formation history.  This result suggests an important change in the
mode of star formation in dense environments over the last few billion
years. Here, we will show that strong \OII\ emission and
\Hd\ absorption is a prominent, perhaps universal, feature of the {\it
composite} galaxy populations of $z \sim 0.5$ clusters, unlike their
present--epoch counterparts.  While the individual spectra of galaxies
may tell us more about the mechanisms that are at work, or the range of
behavior that is present, these composite measurements are a powerful
tool in demonstrating the importance of pure time evolution of cluster
populations versus cosmic variance that may relate to the dynamical
history of individual clusters.

The paper is organized as follows:  In \S2 we describe the
cluster samples and in \S3 we describe the data used in this 
paper and the method of co--adding spectra in clusters and
the field from $z = 0.03$ to $z = 0.91$.  In \S4 we 
discuss the results of comparing intermediate-- and low--redshift
clusters by means of their composite spectra, and the potential
for using such high S/N spectra to answer more detailed questions
about the stellar populations and the history of star formation.
In \S5, we focus on using \Hd\ as a proxy for intermediate--age 
stellar populations.  Specifically, we use the measurements of 
the equivalent widths (EW) of \Hd\ in the composite spectra to 
discuss possibly disparate results found for the Morphs sample 
compared to those published for CNOC1 clusters. In \S6 we
summarize our results.

\section{Cluster samples}

Table 1 lists the clusters and the field samples for which we have
performed the coaddition of spectra. There are eight clusters
from the Morphs intermediate--redshift cluster sample 
(D99),\footnote{The two southern clusters studied at the NTT, 
CL0054 and CL0412 of the D99 study, were not used because they 
contain a high fraction of spectra with poor sky subtraction.} 
supplemented by new data for A851 (Oemler \et\ 2004), and an 
extensive sample of low--redshift clusters studied by Dressler 
\& Shectman (1988, hereinafter DS). The number of coadded spectra 
for each group is listed in Table 1.

These two samples compare well in terms of luminosity depth; 
generally, the cluster samples are magnitude--limited and 
unbiased by type or color. Both are complete,
in the sense that the spectroscopic samples follow the
Schechter luminosity distribution down to the same absolute
magnitude $M_V \sim -19.5$ (see, e.g., Fig. 6a of D99), 
although the low--z sample is more densely sampled by 
factors of 2--3. The completeness of each sample falls rapidly
for fainter galaxies; the small number that are included contribute
less than 10\% of the total light. With respect to area (or volume), 
the DS sample clusters were studied within a radius of $R \sim 2$ 
Mpc,\footnote{We adopt a standark $\Lambda$ cosmology
with a Hubble constant $H_{o} = 70~\kms Mpc^{-1}$ throughout this
paper.} while the Morphs clusters have been studied over 
a somewhat smaller $R \sim 1.5$ Mpc radius.  This is a small 
difference compared to the core radii, which at any rate might be 
expected to result in an undersampling of star--forming galaxies 
in the distant clusters.

All of the Morphs clusters are rich, as confirmed by our spectroscopy
and the large number of {\it bona fide} cluster members. None of them
is contaminated by other significant structures along the line of sight
(see Fig.~2 in D99), a common motivation for choosing clusters by X--ray
luminosity. Interestingly, the Morphs clusters span a wide range in
X--ray and optical properties (e.g. Smail et al.\ 1997).  The structural
properties of these clusters are diverse, including both regular, very
concentrated clusters and irregular ones with clear substructure, such
as Abell 851.  Two of the Morphs clusters, CL0016+1609 and 3C295, are
very luminous ($\rm L_X \sim 12 \times 10^{44} \rm~and~L_X \sim 6.4
\times 10^{44} \, ergs \, s^{-1}$, in the 0.3--3.5 keV range). For
comparison, Coma has an X--ray luminosity intermediate between these
two. The other Morphs clusters have progressively lower X--ray fluxes;
overall they span a factor of 17 in X--ray luminosity.  Abell 851, a
rich but highly substructured cluster (Oemler \et\ 2004), has $\rm L_X
\sim 2~10^{44} \, ergs \, s^{-1}$ (see also De Filippis, Schindler, \&
Castillo-Morales 2003).  (These X-ray luminosities are from 
heterogeneous sources, as described in Smail \et\, Section 2; here
we are concerned only with an approximate ordering and range.)

The DS sample of nearby clusters spans an even larger range in X--ray
luminosities than the Morphs sample (a factor $\sim 24$ in $L_X$).
Abell 754 has an X--ray luminosity comparable to the brightest Morphs
cluster, CL0016+1609, but overall, the low--redshift sample contains
more lower mass, lower X--ray luminosity clusters than the Morphs sample
(Ebeling \et 1996, Girardi \et 1998).

In order to get some indication of behavior at even higher redshift we
have used data from an unpublished study of four clusters $0.65 < z <
0.76$ by Dressler and Gunn (referred to as DGhiz) --- this sample is
less deep in terms of absolute magnitude and less uniformly sampled.
The DGhiz sample consists of 60 cluster members in CL0020+0407 ($z =
0.697$), CL0231+0032 ($z = 0.740$), CL1322+3027 ($z = 0.757$),
CL1322+3114 ($z = 0.696$) from the Gunn, Hoessel, \& Oke (1986)
catalog.  Data for another such cluster, MS1054 at $z=0.83$, has been
generously contributed by van Dokkum and collborators.  These higher
redshift samples are used for illustrative purposes only and no
conclusions in this study are based on them.

\section{Spectral data and technique}

The DS spectra were sampled with 2 x 4 arcsec apertures, a considerably
smaller physical scale ($\sim 2 \times 4$ kpc) than the 1.5 arcsec
slits with typical length of 3.0 arcsec ($\sim 9 \times 18$ kpc) used
in the Morphs study. This difference is exaggerated since the
light is more concentrated than these latter dimensions would suggest,
nevertheless, this mismatch is potentially a matter of concern. The 
situation of robust star formation surrounding a very old bulge 
component is quite rare in present--epoch galaxies (and perhaps at
higher redshift as well), and early--types rarely have strong population
gradients of any kind, nevertheless, this is one area where a better
match in the sampling could and should be made.  The sense of the bias
here would be to underestimate star formation in the present--epoch
clusters.

Each of these studies has generated a {\em field} sample as well.
The DS field sample ($z < 0.06$) is as representative as the DS 
cluster sample, but the Morphs field sample ($0.35 < z < 0.6$)
is likely to be more heterogeneous than the Morphs cluster sample 
because of the way the wide redshift range plays off against
the luminosity function.  Although this is not a serious 
problem for this paper, it means that any conclusions based 
on a comparison of cluster and field at a given redshift are 
tentative. More complete field surveys and additional 
cluster samples do exist; they would be welcome additions to 
these analyses.

The DGhiz study adds another intermediate--redshift field 
sample which should be comparable in quality to the Morphs
field sample, as well as a higher--$z$ ($0.65 < z < 0.95$) field 
sample which is considerably smaller and almost certainly subject 
to serious selection effects. We judge the latter to be
too ill--defined to be used in the present study.

We have excluded identifiable AGNs from our sums, using the
criteria of D99, which amounted to 1 DS galaxy, 7 in the Morphs 
cluster and field galaxies, but no DGhiz galaxies.  For the Morphs 
cluster sample we have chosen only spectra of quality 1--3 (see D99); 
we have used any and all spectra from the DS study (all of which have 
high S/N compared to the distant sample) and all spectra with measured 
redshifts (60 cluster members) from the DGhiz sample.

Ideally, one would combine all the light from all galaxies to form a
luminosity--weighted composite spectrum.  Of course, this is impossible
from the start because slit spectroscopy samples only a portion of the
light in each galaxy.  As an alternative, {\em for the cluster samples}
we have normalized each spectrum to unity over the interval 4200-4600\AA\
(approximating the rest-frame R-band but avoiding the Balmer
absorption and [O II] emission) and weighted its contribution
by the R-band galaxy luminosity, (The field samples are sufficiently
inhomogeneous that we don't consider such a weighting to be useful.)
That is, we have assumed that each spectrum has sampled a
representative piece of that galaxy's stellar population (as also assumed 
in previous galaxy--by--galaxy analyses). Examples of the results of the
summing procedure is shown in Figure 1, which presents the composite
spectra in the \OII\ to \Hd\ wavelength range for 5 of the Morphs
clusters and 5 of the DS low--redshift sample.  The high S/N of these
composite spectra --- typically 5--10 times that of individual spectra
(see Dressler \et\ 1997) --- is apparent.  The MS1054 composite
spectrum, kindly provided by P. van Dokkum, is spectral-flux-weighted
sum (through the slit) of 72 I-selected cluster members (see
van Dokkum \et\ 2000).

As we have done for the individual Morphs spectra, we restrict
ourselves in this study to measuring two spectral features. We use the
strength of \OII\ emission as an indicator of recent star formation;
H$\alpha$ would be preferable, but this is not generally covered by the
spectra.  To isolate the intermediate--age population associated with
starbursts or decaying star formation we use \Hd\ absorption, whose
properties are discussed in more detail in P99. \Hd\ is a reliable
indicator because it is the strongest of the Balmer lines for which
there is minimal contamination by emission.  $H\gamma$ and $H\beta$
lines are intrinsically stronger, but are more filled in by emission in
galaxies with strong star formation (see, e.g., Trager \et\ 2000a,b;
Trager 2004). The higher order Balmer lines are even less subject to
filling, but they are weaker and in less clean regions of the
spectrum.  The shape of the continuum around \Hd\ is also the cleanest
of any of the higher--order lines, being mostly devoid of strong nearby
features, although it is not without difficulties, as discussed below.

%
%
\hbox{~}
\centerline{\psfig{file=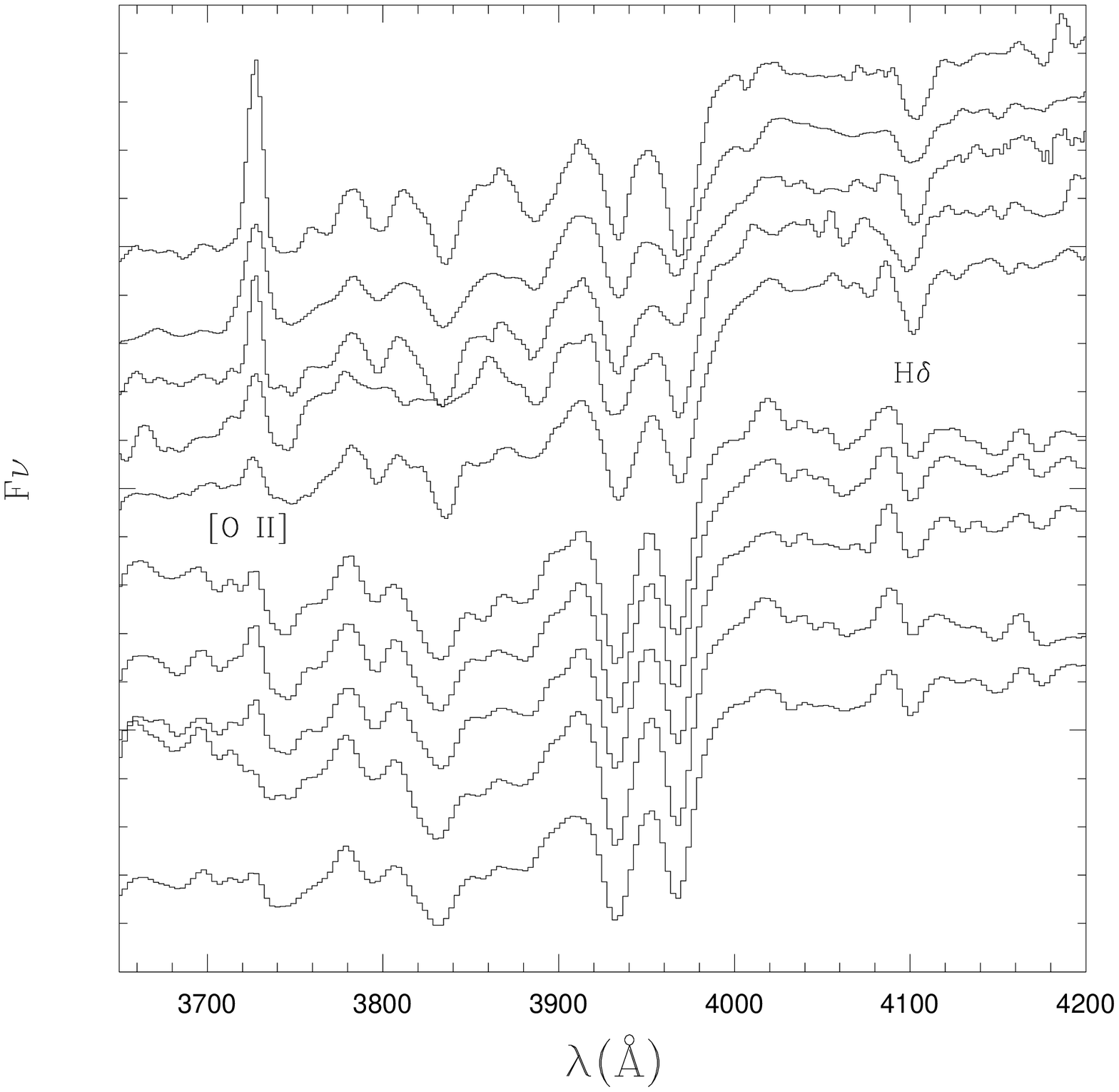,angle=0,width=4.5in}}
\noindent{\scriptsize
\addtolength{\baselineskip}{-3pt} 

\hspace*{0.3cm}Fig.~1.\ Composite spectra for 5 intermediate--redshift
clusters (above) and 5 low--redshift clusters (below) from the Morphs
and DS samples. We have chosen the better examples, but \OII\ and \Hd\
are seen in all clusters of both samples; values for all clusters are
tabulated and included in subsequent plots. The \OII\ and \Hd\ features
are clearly stronger in the spectra of the intermediate--redshift 
clusters.  Also apparent is the stronger contribution in the 
low--redshift cluster spectra of various lines around \Hd\ that 
depress the continuum and leave what appears to be a peak immediately 
to the blue of H$\delta$, as described in Section 4.2.  On this and
in the following plots, the ordinate F$\nu$ in ergs s$^{-1}$ hz$^{-1}$
is in relative units.

\vspace*{0.15cm}

\addtolength{\baselineskip}{3pt}
}

The DS data have a spectral resolution of FWHM $\sim5$\AA, while
the Morphs spectra have a lower resolution of FWHM $\sim9$\AA, in
the rest frame.  In order to correct for this difference, the DS
data were smoothed by a Gaussian kernel to lower their spectral
resolution to $\sim9$\AA\ FWHM.  Also, to approximately
match the 2.8\AA\ per pixel binning (rest frame) of the Morphs
spectra, the DS data were rebinned to 3.0\AA\ per pixel.
We found no strong sensitivity of our measurements to such
adjustments, but differences were measured at the 10--20\% level,
primarily for \Hd, which tends to drop in equivalent width with
lower spectral resolution.  Likewise, there should be a tendency 
to underestimate \Hd\ strengths for galaxies with lower velocity
dispersion, $\sigma < 150~\kms$ in the DS spectra and $\sigma 
< 200~\kms$ in the Morphs spectra.  This suggests that the
effect, if significant, goes in the direction of diluting the 
true strengths  of \Hd\ lines in the distant sample.

The measurement of \OII\ emission is straightforward as this feature is
unresolved in our data and found on a nearly--flat continuum.  We use a
line--fitting technique, as described in D99, however, a bandpass
technique (subtracting the level of a band containing the feature from
the average of two straddling continuum bands) would give as good a
result for [O II]. The line--fitting technique is preferred for \Hd\
because, as we discussed in Section 4.2, the line width and the
character of the surrounding continuum change with the strength of the
feature and the character of the stellar ``background,'' which itself
seems to change with redshift.  The high S/N of the composite spectra
guarantees excellent fits to both \Hd\ and \OII, with typical errors in
equivalent width of less 10\%.

\section{Comparing clusters through composite spectra}

The measured line strengths, \EWOII\ and \EWHd\ are given in Table 1.
The results of this first attempt to study composite cluster
populations, shown in Figure 2, are remarkably clear.  The
intermediate--redshift clusters all average \Hd\ strengths of
$\sim$2.0\AA\ and have substantial \OII\ emission varying between
-3\AA\ and -10\AA\ equivalent width; both are indicative of
substantial ongoing star formation. In contrast, each of the
low--redshift clusters shows much weaker \EWHd\ $\sim$ 1.3\AA --- this
is not just 35\% weaker, but typical of a population that has had {\em
no significant star formation} for several Gyr --- and much weaker
\EWOII\ = $-$2\AA\ to $-$5\AA. This difference in spectral
characteristics, in particular, the appearance of many systems with
strong Balmer lines in intermediate--redshift cluster samples noted by
Dressler \& Gunn (1983, see also Couch \& Sharples 1987), appears to
be primarily dependent on epoch. There may be some real sensitivity to
cluster properties buried in the scatter in Figure 2, but much more
data on each cluster will be needed to be certain that these are not
just statistical errors.

It is interesting to note that, unlike the strong evolution in these
features from low to intermediate redshift, there is no further
significant increase for the even higher--redshift DG sample or for
MS1054--03 at $z = 0.83$ (derived from data kindly provided P. van
Dokkum).  These clusters have composite values of \EWHd\ and \EWOII\
that are consistent within the errors with the Morphs
intermediate--redshift sample.  Although these first indications
suggest that \EWHd\ and \EWOII\ do not continue to increase at higher
redshift, it is clearly too early to tell.  A particular concern is
that the higher redshift clusters are not sampled as deeply, which
means that fainter galaxies, which are by--in--large more likely to be
the places of higher rates of star formation, comprise a smaller
fraction of the sample.

%
%
\hbox{~}
\centerline{\psfig{file=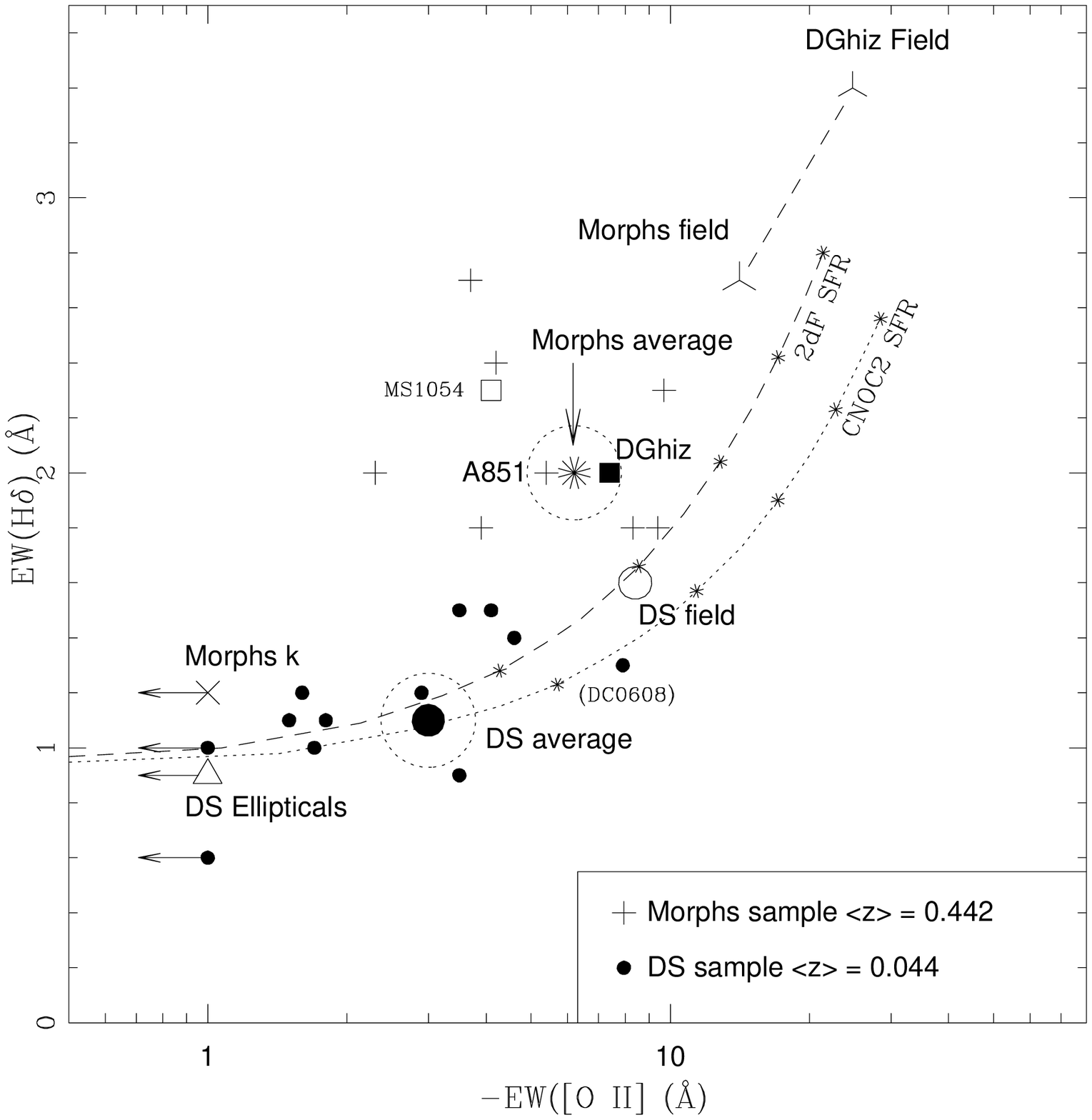,angle=0,width=5.5in}}
\noindent{\scriptsize
\addtolength{\baselineskip}{-3pt} 

\hspace*{0.3cm}Fig.~2.\ Measurements of \EWOII\ and \EWHd\ for the
various field and cluster samples of Table 1.  There is a clear
separation of the ``mean'' values of \Hd\ for the Morphs and the DS
samples, but there is also no overlap in the individual measurements
for the clusters studied. Even higher--redshift clusters may share the
properties of the Morphs sample, with no obvious further increase in
activity from $z \sim 0.5$ to $z \sim 0.8$.  As discussed in \S4.1, the
two lines represent different mixes of passive galaxies and
continuously star--forming galaxies with \OII\ distributions from the
2dF Survey (dashed line) and the CNOC survey (dotted line). The small
asterisks along the curve mark the appropriate values of \EWOII\ and
\EWHd\ for 20\%, 40\%, 60\%, 80\%, and 100\% mix of galaxies undergoing
continuous star formation.  Although such mixes adequately describe the
low--redshift DS clusters and field, they fail to match any of the
higher--redshift cluster or field samples, indicating the importance of
starbursts to these galaxy populations.

\vspace*{0.2cm}

\addtolength{\baselineskip}{3pt}
}

Just as interesting and much more certain is the lack of overlap in
the composite \Hd\ strengths for the two primary samples: although the \OII\
strengths overlap considerably, there are no present--epoch clusters
with strong \Hd\ and no intermediate--redshift clusters in which it is
weak.  As we discuss below, there is the suggestion by Balogh \et\
(1999) and Ellingson \et\ (2001) that strong X--ray--emitting clusters at
intermediate redshifts show less star formation activity than the
Morphs clusters, but the spectroscopic data presented here certainly
offer no examples of the inverse, that is, present--epoch clusters
resembling those of the Morphs sample.  The universality of the result
in Figure 2 must await further data from other clusters, but there
seems little doubt that clusters with strong \Hd\ absorption in their
composite spectra are common at $z\sim 0.5$, and at best rare today.

In as much as the composite spectra of the 8 individual Morphs
clusters are very similar, we produce the sum of the galaxies in {\it
all} 8 clusters in Figure 3 (also plotted as a point in Figure 2).  
By subdividing this grand--composite spectrum into the Morphs spectral
classes (described in D99), we create a recipe for a $z \sim 0.5$ 
cluster, shown in Figure 3 as the percentages of each type of 
spectrum that make up the whole.  Even though $\sim$70\% of the 
galaxies do not have strong Balmer lines (primarily k and e(c) 
types) the signature of the \Hd\ line is obvious in the composite 
spectrum --- this shows how strong is the contribution of \Hd\ from 
the remaining $\sim$30\%, a fact that we attribute largely to 
starbursts, as we discuss in the next section.

%
%
\hbox{~}
\centerline{\psfig{file=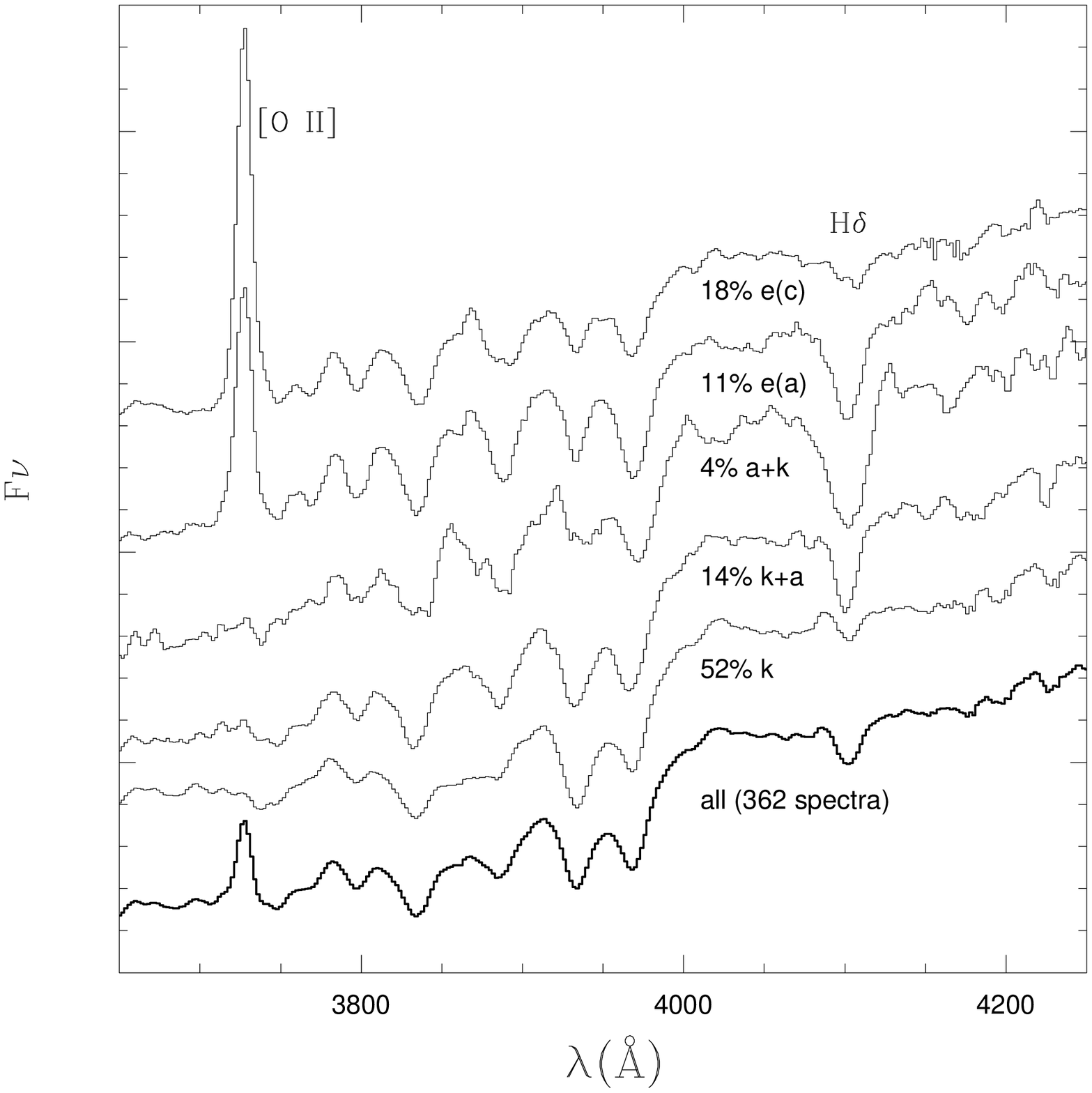,angle=0,width=4.5in}}
\noindent{\scriptsize
\addtolength{\baselineskip}{-3pt} 

\hspace*{0.3cm}Fig.~3.\ The grand composite spectrum for galaxies
in all 8 clusters at $z \sim 0.5$, excluding 7 AGN's and 10
others with poor sky subtraction. The contribution to the combined 
spectrum (bottom) of \OII\ and \Hd\ from the "active" galaxies, 
k+a, a+k, and e(a), is clearly seen.  The fraction of each type 
is noted.

\vspace*{0.2cm}

\addtolength{\baselineskip}{3pt}
}

The comparison with field samples is also interesting.  As has been
found in other field surveys, the field at intermediate--redshift shows
a considerably stronger \OII\ than at $z \sim 0$, indicating a higher
rate of star formation in the relatively recent past (e.g.,
Broadhurst, Ellis, \& Shanks 1988).  Not previously demonstrated,
however, is the significantly greater strength of \Hd\ in field
galaxies at $z \sim 0.5$ and $z \sim 0.75$.  Our two field samples
at $z \sim 0.5$, from the Morphs study and the DGhiz sample, 
both exhibit a much stronger \Hd\ 
absorption than the low--redshift DS field.  The obvious question is, 
for both cluster and field, do these stronger \Hd\ lines represent 
simply a higher fraction of galaxies with strong, continuous star 
formation, or do they also point to a significant increase in the 
fraction of  starburst galaxies in the field?  To answer this 
question, we must compare to the expected strength of \Hd\ for 
populations of galaxies undergoing continuous star formation. 

\subsection{Expectations for continuously star--forming galaxies}

In this section we determine the mean values of \EWOII\ and \EWHd\ 
that are expected for different mixes of passive galaxies and
continuously star forming galaxies.  Our goal is to test 
whether any such populations, regardless of proportions,  
are able to account for the locus of points we find for the 
intermediate--redshift clusters and field.

It might be possible to calculate the expected mean values of \EWOII\ 
vs \EWHd\ using stellar population models, but we prefer to stick as
closely as possible to empirical data.  The present--epoch field
population includes only a very small fraction of starbursting systems
(see, e.g., Zabludoff 1996), so we assume that this population can
represent the universe of galaxies undergoing continuous star formation
(or no star formation).  Furthermore, we assume that, whatever the
scatter in the relationship \EWHd\ = $f$(\EWOII) for continuously
star--forming galaxies, there is a well-defined mean of \EWHd\ for each
\EWOII\ that does not vary significantly with epoch or environment, at
least since $z = 1.0$.\footnote{For young populations, $\tau < 1
Gyr$, the continuum in the \OII\ to \Hd\ part of the spectrum comes from
the young stars themselves, so it will not evolve with epoch.  For old
populations, $\tau > 5$ Gyr, the continuum slope will slowly change
with time, but this will be a very small effect in an era of declining
SFR.  Only for intermediate age populations might there be a measurable
difference in the relationship of \EWOII and \EWHd, but with the
proximity in wavelength of these features, the effect should be second
order at best.} If that is so, then the expected mean values of
\EWOII\ and \EWHd\ of any population depend on only two independent
relations: the \EWHd\ = $f$(\EWOII) relation, and the distribution of
\EWOII\ values within the population.

We determine the mean \EWHd\ = $f$(\EWOII) relation for continuous
star--forming galaxies in two steps.  Goto (2003) has 
determined values of \EWOII\ and \EWHd\ for 95,479 galaxies in the
Sloan Digital Sky Survey, using line fitting methods similar to those
which we have employed. These galaxies are overwhelmingly
low--redshift field galaxies, and thus represent the population we
wish to use. Goto's published data on \EWHd\ have been corrected for
infilling of the absorption line by emission. Our
observations are of insufficient resolution and S/N to do the same,
but Dr. Goto has kindly provided to us the uncorrected values, which
we use. Furthermore, although Goto's method is very similar to ours,
it is not identical. To correct for systematic differences in \Hd\
linestrengths between Goto's and our techniques, we use the more
limited measurements of \EWOII\ and \EWHd\ in the DS field population,
to apply a shift of zero point and slope to the Goto relation. The
resultant curve, which we shall take as the expected relation between
\EWOII\ and \EWHd\ for continuously star forming galaxies in our
measurement system, is presented in Figure 4.

%
%
\vspace*{0.1cm}
\hbox{~}
\centerline{\psfig{file=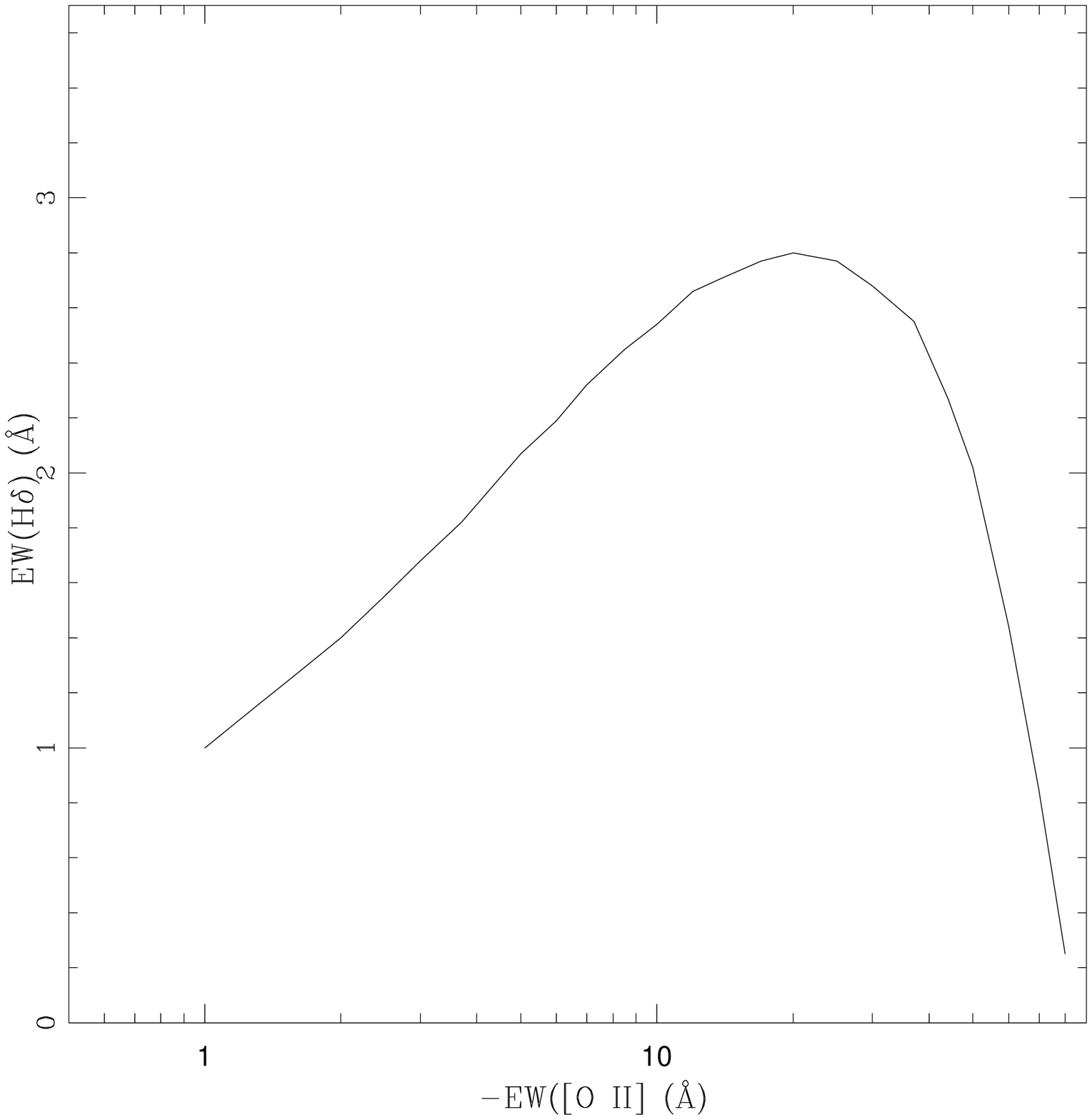,angle=0,width=3.5in}}
\noindent{\scriptsize
\addtolength{\baselineskip}{-3pt} 

\hspace*{0.3cm}Fig.~4.\ Adopted relationship between mean \EWHd\ and
mean \EWOII\ for galaxies undergoing continuous star formation.

\vspace*{0.2cm}

\addtolength{\baselineskip}{3pt}
}

Figure 4 shows that, due to infilling, and the fact that \EWHd\ peaks
for A stars but then decreases for younger populations that include O
and B stars, there is a well--bounded range of \Hd\ strengths for
continuously star--forming systems.  However, this range is more than
large enough to span the measurements of \EWHd\ in our composite
samples of cluster and field galaxies, as seen in Figure 2.  This
suggests that populations of continuously star--forming galaxies {\em
could} reproduce most of the line strengths we observe. In fact, as we
now demonstrate, they do not.

Balogh \et (2004) have shown, using the 2dF galaxy sample,
that star formation is remarkably uniform in the sense that, although 
the mean star formation rate (SFR) varies radically with environment, 
this is due almost entirely to variation in the {\it fraction} of
star--forming galaxies.  {\em Among only star--forming galaxies, 
the distribution of star formation rates is nearly independent of 
environment}. Wilman \et\ (2004) have demonstrated the same phenomenon 
at $z \sim 0.4$. Indeed, even the variation of the SFR distribution 
of star-forming galaxies with epoch is modest: at higher redshifts 
it is skewed somewhat towards higher rates, but the mean \EWOII\ 
for star--forming galaxies at $z \sim 0.4$ is only about 20\% higher 
than it is today (see Bower \& Balogh 2004).

As a consequence of the small variation in the distribution of SFRs of
star--forming galaxies {\em and} the small range in \EWHd\ with
\EWOII\ shown in Fig. 4, the expected mean \Hd\ strength of
populations of continuously star--forming galaxies is a very stable
number. We calculate that number for two cases: a low SFR case taken
from the low redshift 2dF group sample, and a high SFR case taken from
the $z \sim 0.4$ CNOC2 field sample. For both, we start with the
observed distribution of \EWOII\ for star forming galaxies, from Wilman
\et\ (2004).  In both cases we extend the distributions to weaker
\EWOII, from the 5 \AA\ limits reported by Wilman \et, down to 1 \AA,
using the data on the LCRS sample from Hashimoto \et\ (1998). For the
low and high SFR samples, which have mean \EWOII\ of 21 \AA\ and 28
\AA, respectively, we calculate mean values of \EWHd\ of 2.8 \AA\ and
2.6 \AA, using the \EWHd\ = $f$(\EWOII) relation in Figure 4.  It is
striking that the mean \EWHd\ is {\em lower} in the population with
higher star--formation, because of the larger fraction of galaxies with
emission infilling and the shift to younger, hotter stellar
populations. For the passive, non--star--forming galaxies, we use the
sum of all E galaxies in the DS sample to obtain mean a value of \EWHd\
$= 0.9$ \AA\ and a limit \EWOII\ $>$ $-1.0$ \AA.

With each of these two sets of mean \EWOII\ and \EWHd\ strengths, we
can now form a one--parameter family of galaxy populations, where the
parameter is the fraction of galaxies with star formation.  The dashed
and dotted lines of Figure 2 show the locus of points of different
mixes of passive and star forming galaxies, for the low and high SFR
distributions.  Since the 2dF group sample and the CNOC field sample
seem to span the range of \EWOII\ distributions, at least out to $z
\sim 0.5$, the region between these two curves in Figure 2 represents
the locus of possible values in this plane for populations of galaxies
with (mainly) continuous star formation.  It is noteworthy that --- as
it should be --- all of the low--redshift populations, both individual
clusters and mean cluster and field populations, are consistent with
this region.  In other words, they are well described by mixes of
passive galaxies and continuously star forming galaxies. In contrast,
all higher--redshift populations lie {\it above}, implying that \Hd\ is
stronger than can be produced from what is today's ``normal'' star
formation. The straightforward conclusion is that starbursts make up a
substantially greater part of the population of both cluster and field
galaxies at higher redshift. This is arguably the strongest conclusion
to be drawn from this paper, that neither the cluster nor field
population at intermediate redshift can be made from a common mix of
today's galaxies: a much higher fraction of galaxies experiencing
starbursts must be invoked.\footnote{We cannot demonstrate from these
data alone that these higher points could not be the result of simply
truncated, rather than bursting, star formation.  The \EWHd\
distributions themselves, rather than their composites, form the basis
for this assertion, as discussed in P99.}

\subsection{More detailed information on stellar populations
from high S/N profiles of \Hd\ absorption}

With the higher signal--to--noise and higher spectral resolution now
available with 6--10 meter telescopes and efficient spectrographs, it
is possible to measure the strengths of individual absorption lines in
the blue region of intermediate-- and high--redshift cluster galaxies.
For example, with Keck--LRIS spectra with a typical
$\mathrm{S/N}\sim60$ per resolution element, and (rest--frame)
resolution of $\sim5$ \AA\ for Abell 851, Trager, Dressler \& Faber
(2004) are able to measure all of the available Lick/IDS absorption
line indices (Faber \et\ 1985; Worthey \et\ 1994).  Some of the Rose
(1985; 1994) blue indices are also available (bright sky lines make
some of these indices unmeasurable).  Indices such as
CN$_2$ (cf. Kelson \et\ 2001), Fe4383, the G band, and Ca
H+H$\epsilon$/Ca K, and even features such as Mn and \ion{Sr}{2} around
4050 \AA, can be used along with higher--order Balmer lines to break
the age--metallicity degeneracy (cf.\ Trager et al., in prep.; Calwell,
Rose \& Concannon 2003) and even study individual abundance ratios
(e.g., Trager \et\ 2000a,b).

Co--adding of spectra will greatly aid these kinds of studies
because it allows the collection of larger, and fainter samples
in equivalent time, albeit at the loss of some specificity as
galaxies of a given morphological type, absolute luminosity,
location, etc.  For example, we
formed composite spectra of 46 E and 24 S0 galaxies in
the 8--cluster Morphs sample, selected only by morphology and
the absence of \OII\ emission.  The result, shown in Figure 5, is 
much stronger \Hd\ in the S0 population than in the
E population.  This suggests that the S0 population, on
average, has been the site of more recent star formation than 
the E population, consistent with the picture presented by
Dressler \et\ (1997) that the S0 population is growing substantially
at the $z \sim 0.5$ epoch.  Note that this result 
contradicts the conclusions of Jones, Smail, \& Couch (2000) who 
used some of the same data used in this study to conclude that the
star formation histories of elliptical and S0 galaxies were
indistinguishable.  Our sample, which has been carefully
culled to remove galaxies with ongoing star formation, seems
to show something very different.  Possible differences in the
two studies are the inclusion of lower redshift clusters, 
clusters that are more evolved dynamically, and a greater
proportion of brighter galaxies in the Jones \et\ study.
%
%
\hbox{~}
\centerline{\psfig{file=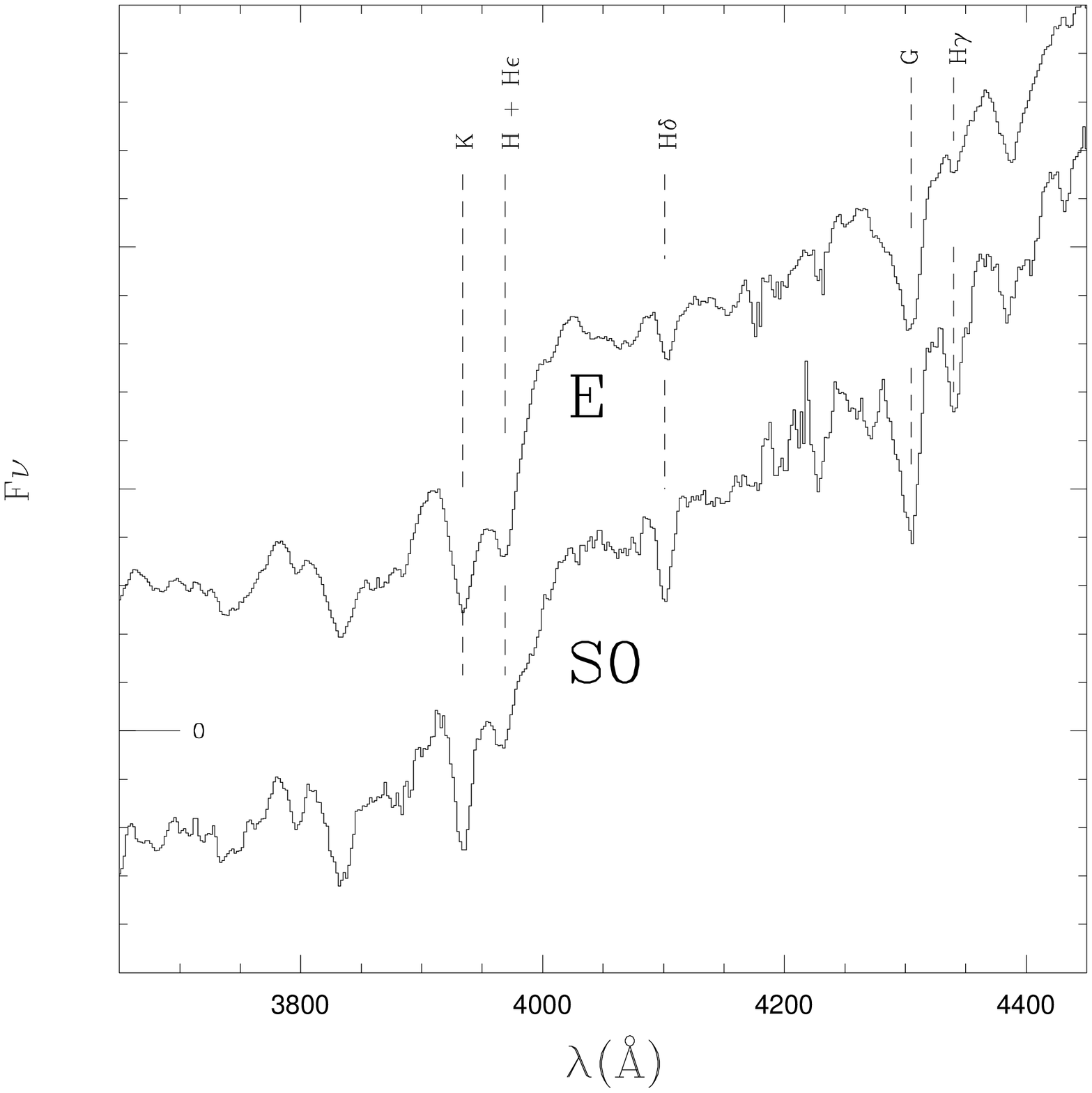,angle=0,width=3.5in}}
\noindent{\scriptsize
\addtolength{\baselineskip}{-3pt} 

\hspace*{0.3cm}Fig.~5.\ The composite spectra for 46 E and
24 S0 galaxies, those in the Morphs 8--cluster sample without
detectable \OII\ emission \EWOII\ $\gs$ 3\AA.  The detection
of stronger \Hd\ absorption in the S0 galaxies supports the
notion that these have had more recent star formation than
the elliptical galaxies.

\vspace*{0.2cm}

\addtolength{\baselineskip}{3pt}
}

In this paper our focus has been on the measurement of \Hd\ with high
S/N composite spectra.  A careful look at its adjacent continuum
reveals an epoch--dependent effect that bears on the question of how
best to measure \Hd\ in spectra of resolution 5--15 \AA\ that is
typical of studies of distant galaxies.
  
The higher S/N composite spectra of Figure 1 clearly show a difference
in the shape of the continuum around \Hd\ in the low--redshift cluster
galaxies compared to those in the distant sample.  There is a 
prominent peak on the blue side of the \Hd\ line of low--redshift
spectra that is not seen at higher redshift.  To provide some
insight into this feature, we show in Figure 6 stellar models 
of an aging, metal--rich stellar population from Vazdekis (1999).
The models exhibit a depression of the integrated stellar continuum 
by the increasing prominence of CH and CN lines.  The exact cause of
this increase is poorly understood at present, but appears to come
from the increased importance of cool giants in old, metal--rich
populations (cf.\ Schiavon et al.\ 2002).

These theoretical models are in qualitative agreement with the
composite data of all the DS (low--redshift) clusters, also shown in
Figure 6. It seems that, at this spectral resolution, the peak to the
blue of \Hd\ is actually a window to the true continuum, which is
depressed by a blend of lines around and over \Hd.  The models confirm,
however, that \Hd\ itself is weak, \EWHd\ = 1.0~\AA\ as measured by our
technique, for a 12 Gyr, solar--metal--abundance population, consistent
with what we have measured. Although the models and observations
roughly agree, it is apparent that this blue peak is even more
prominent than predicted even by 15 Gyr--old, [Fe/H] = +0.2 model.  The
change between our 5--Gyr--old Morphs sample, for which the feature is
not seen, and the present--epoch population where it is seen so clearly,
indicates a more rapid evolution than the model in Figure 6 predicts.
We plan in a future paper to investigate the implications of changing
contamination of \Hd\ with epoch, which is relevant not only to actually
measuring line strengths but more importantly to exploit the diagnostic
value of the weaker features as tracers of the stellar population
history.

%
%
\begin{figure*}
\centerline{\psfig{file=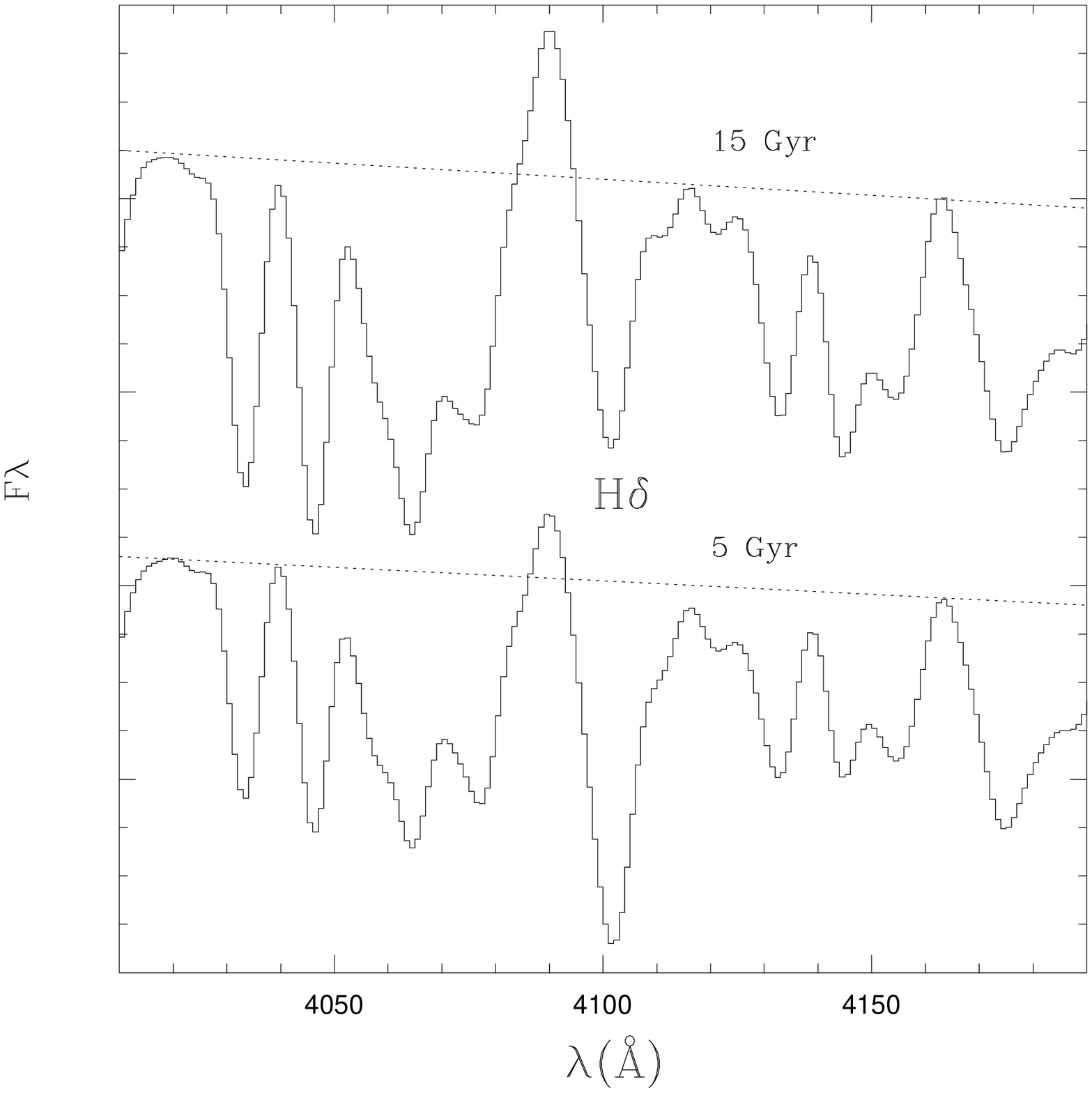,angle=0,width=3.0in}\hspace*{0.5cm}
\psfig{file=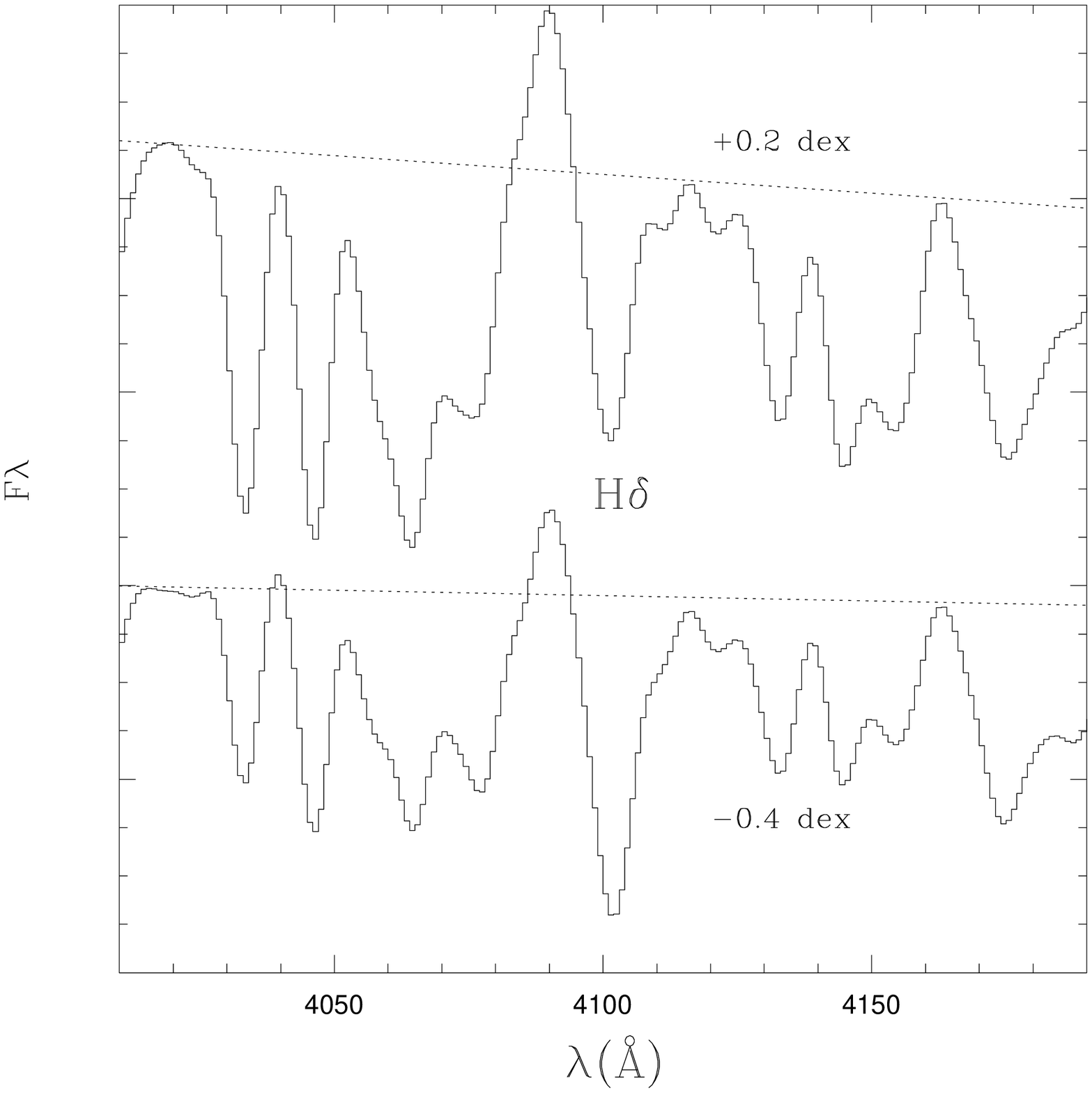,angle=0,width=3.0in}}
\centerline{\hbox{\psfig{file=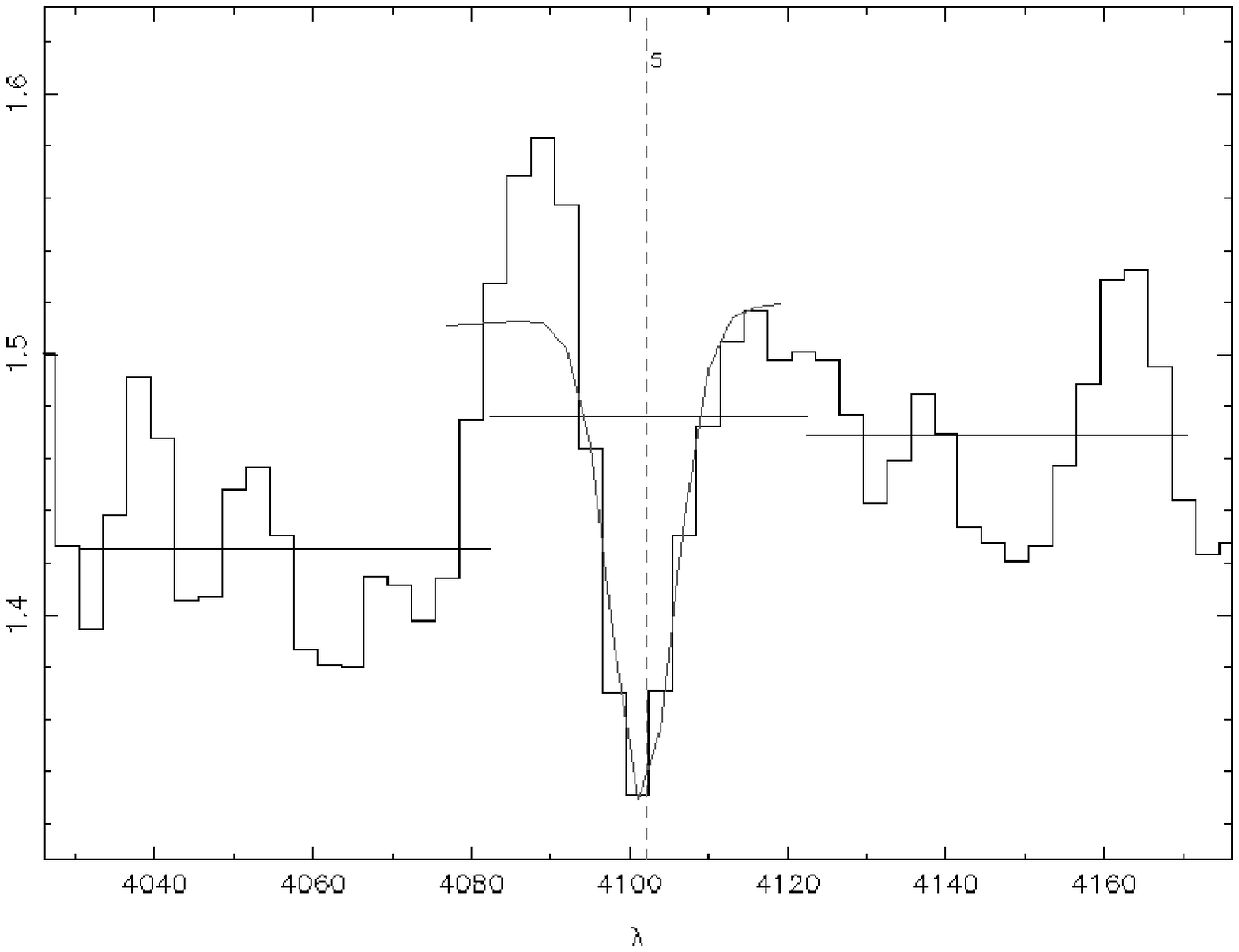,angle=0,width=4.0in}}}

\noindent{\scriptsize
\addtolength{\baselineskip}{-3pt} 
\hspace*{0.3cm}

\hspace*{0.3cm}Fig.~6.\ Models by Vazdekis (1999) of the evolution of
the integrated stellar continuum around \Hd.  The ordinate is
in relative flux units ergs s$^{-1}$ \AA$^{-1}$.  The peak to the blue of
\Hd\ is apparently clear continuum in a growing blanket of absorption.
a) (top left) the change as a stellar population ages: a 15 Gyr--old
population (top) as compared to one at 5 Gyr.  For models of ages 5,
8, 12, and 15 Gyr, we measure \EWHd\ of 1.4, 1.3, 1.0, and 0.9 \AA,
respectively.  b) (top right) a stellar population of age 12 Gyr at two
different metal abundances: $+$0.2 dex (top) compared to $-$0.4 dex. For
models [Fe/H] = $-$0.4, $+$0.0, and $+$0.2 dex we measure \EWHd\ of 1.3,
1.0, and 1.0 \AA, respectively.  The effect we see of larger
blanketing in the \Hd\ continuum could be the result of aging or
increasing metal abundance, or both. An old and/or metal rich
population has a much stronger blanketing in this region, resulting in
the significant peak to the blue of \Hd\ where the blanketing is
fortuitously low. This complicates the measurement of \Hd\ strength,
particularly in old, metal--rich, stellar systems.  c) (bottom)
Composite spectrum of 12 DS clusters at the present epoch, showing the
feature predicted in the models, at least in a qualitative sense.  The
line--fitting technique is able to make a sensible measurement of \Hd,
while the bandpass method yields a negative value, as shown by the
levels marked in the figure.

\vspace*{0.2cm}

\addtolength{\baselineskip}{3pt}
}

\end{figure*}

A final example of the information carried by the \Hd\ line is shown in
Figure 7, where we extract the \Hd\ line for all k+a, e(a), and a+k
galaxies in the Morphs sample.  Comparing the spectra from k+a, e(a),
and a+k shows a progression of broader (stronger) \Hd\ with growing
asymmetry.  This is the signature of an earlier population of A--stars;
late--B and early--A stars have Stark--broadened Balmer lines that are
much stronger than even a typical A5V star. The a+k spectrum in
particular appears to show a large contribution from stars as early as
A0, which fixes the time of the burst to 300 $\pm$ 100 Myr, while the
composite k+a spectrum represents a typical age of 1 Gyr or more.  
This agrees with the chronology we have proposed in P99 for
the population of starburst galaxies and the different spectral
types we have identified (see also Barger \et\ 1997).

%
%
\begin{figure*}
\vspace*{0.2cm}
\centerline{\hbox{\psfig{file=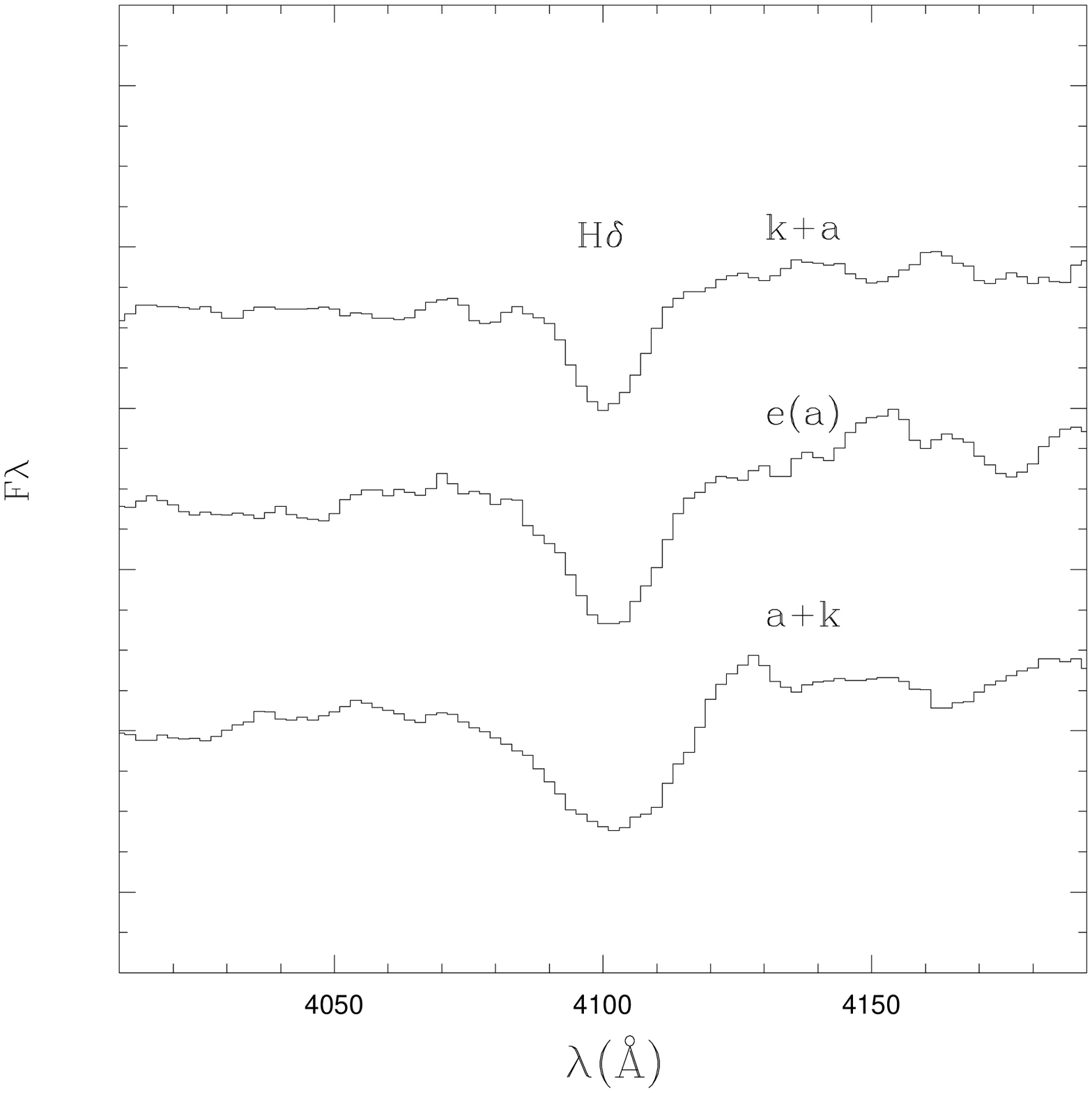,angle=0,width=3.0in}}}
\vspace*{0.2cm}
\centerline{\hbox{\psfig{file=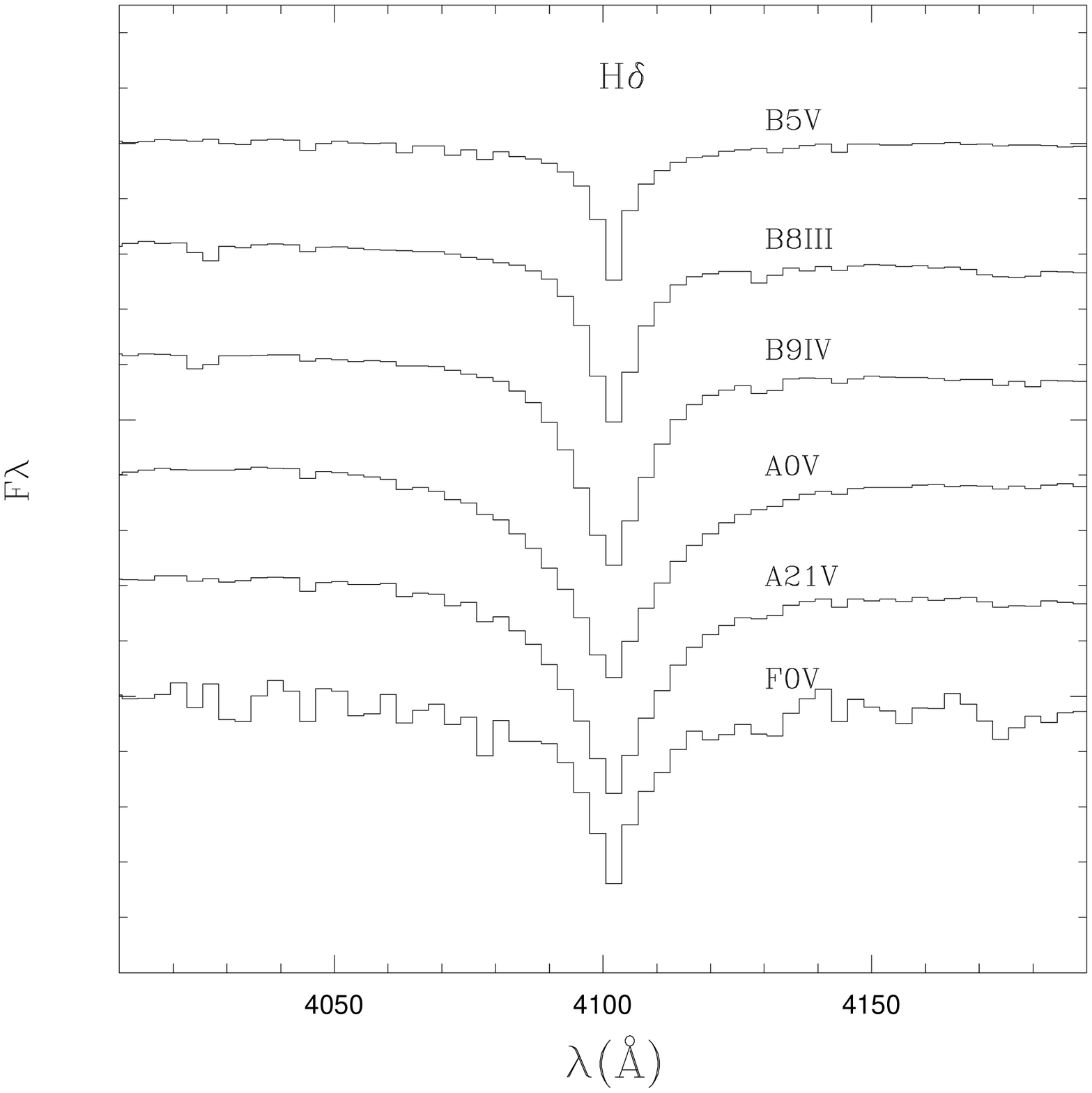,angle=0,width=3.0in}}}

\noindent{\scriptsize
\addtolength{\baselineskip}{-3pt} 

\hspace*{0.3cm}Fig.~7.\ Change in line profile with increasing
\Hd\ strength. a) (top), composite k+a spectra, as defined in
Poggianti \et.; b) (middle) composite e(a) spectra; c) (bottom)
composite a+k spectra.  The growing width and asymmetry with 
increasing \Hd\ strength suggests Stark--broadened profiles of 
stars around type A0 in contrast to either earlier or later types.
This pegs the time of the burst to 300 $\pm$ 100 Myr, while the
composite k+a spectrum is typical of a burst with an age $\sim$1 Gyr.  

\vspace*{0.2cm}

\addtolength{\baselineskip}{3pt}
}

\end{figure*}

\subsection{Measuring \Hd\ by line--fitting or bandpasses?}

The two effects regarding \Hd\ just discussed --- that the continuum
changes with metalicity and age, and that the \Hd\ line becomes broad
and asymmetric in young $\sim10^{8}$ year--old populations --- are
problematical for the measurement of \Hd\ with the commonly--used 
bandpass technique. As \Hd\ grows in strength its width also
increases substantially, as can be seen in Figure 5.  Our
line--fitting technique and bandpass measurements, using the CNOC1
definition (Balogh \et\ 1999) yield nearly equivalent results
between 2\AA\ $<$ \Hd\ $<$ 6\AA, but begin to depart for stronger \Hd.
For the a+k spectrum shown in Figure 7c, the bandpass method
underestimates the 8.5~\AA\ equivalent width of \Hd\ by 15\%.

The measurement of {\it weak} \Hd\ lines with the bandpass method is
much more problematical. The continuum features around \Hd\ are a function
of age and metalicity, as we discuss above. This presents a problem in
particular for measurements of \Hd\ in present--epoch galaxies with
predominantly old stellar populations.  For example, the composite
spectrum for all clusters in the DS sample (Fig. 1) has an \EWHd\
$= 1.3$\AA\ through the line--fitting technique, but
the bandpass definition used by Balogh \et\ (1999) results in a
negative equivalent width ($-$0.8\AA) due to absorption features that
depress the straddling continuum bands (see Figure 6c).

A disadvantage of the line--fitting technique the Morphs have used is
that it requires relatively good spectra in order to constrain the
free parameters of fitting the continuum and the line shape.  The
bandpass method is robust as a function of signal--to--noise, but has
the problems just discussed.  We conclude that, if spectra are of
sufficient quality (eg., Morphs spectra quality 1--3), including
of course the high S/N composite spectra used here, the
line--fitting technique is preferred.

Goto (2003) has reached similar conclusions in his analysis of the SDSS
spectra.  That project has the additional advantage of higher spectral
resolution, sufficient to resolve Balmer emission in the core of the
Balmer absorption and correct accordingly.  In future studies of
distant galaxy populations, at least for $z < 1$, it may become common
to have adequately high resolution to match the SDSS measurements with
its enormous sample of galaxies, most of them at $z \ls 0.1$, that span
such a wide array of galaxy properties.

\section{Implications for the controversy over starbursts in 
intermediate--redshift clusters}

In this study we have concentrated on measurements of \OII\ emission and
\Hd\ absorption as broadly indicative of star formation histories.  
That galaxies in intermediate redshift clusters commonly have stronger 
\OII\ emission, and higher rates of star formation, than do those in 
clusters today is an accepted fact which few dispute.\footnote{However, 
De Propris \et (2003) have raised the question, based on the smaller
Butcher--Oemler effect found for IR--selected samples, of whether the
masses of these starforming galaxies are far below M* (see, however,
Barger \et\ 1997). This issue is now under study using kinematic 
measurements of intermediate--redshift disk galaxies in and out 
of clusters.}

On the other hand, the incidence of strong \Hd\ absorption, and its
significance for the past history of star formation in these galaxies,
has been contentious.  In past papers (P99, D99) we have
presented data to demonstrate that 20--30\% of the
galaxies in the Morphs clusters have strong to very strong \Hd\
absorption, and have argued that many, perhaps most, of these galaxies
experienced a burst of star formation during their ingress to the
cluster.  If so, this is an essential clue to the processes which
transform galaxies after they enter clusters (see also Tran \et 2003). 

However, the observation of enhanced Balmer absorption has been
questioned by the CNOC1 group (Balogh \et\ 1999, hereinafter B99;
Ellingson \et\ 2001), who have measured \OII\ and \Hd\ strengths for 1823
galaxies in a sample of strong X--ray--emitting clusters at $0.17 < z <
0.55$ (Yee, Ellingson, and Carlberg 1996).  In particular, B99 conclude
that the fraction of strong Balmer--line galaxies (\EWHd\ $> 5$\AA,
\EWOII\ $< -5$\AA) is less than 5\% and, in fact, consistent with 
{\it no increase} compared to the field at comparable redshift or, 
by comparison with the Las Campanas Redshift Survey (Zabludoff \et\ 
1996), no increase compared the present--epoch field.  In an attempt 
to account for the difference, B99 questioned the (1) representative 
nature of the Morphs clusters, (2) the sampling of galaxies within 
those clusters and (3) the reality of the detections of Balmer lines.

We will return to (1) and (2) in the following section.  Concerning
(3), the technique of composite spectra developed in this paper allows
a clear demonstration of the reality of strong \Hd\ absorption in the
Morphs clusters. An inspection of Figure 1 shows the prominence of
\Hd\ absorption (all the more remarkable since the majority of cluster
galaxies, with their classical K--type spectra, have very weak Balmer
lines), and the contrast with the DS sample of present--epoch clusters
is striking.  

However, as B99 have pointed out, relying on a single feature such as
\Hd\ can be risky in noisy data, because co--adding of features that are
pure statistical fluctuations will also produce what appears to be a
more certain detection.  Hence, a great deal of the discussion in B99
focuses on ``correcting'' the CNOC1 data for the noisy measurements of
\Hd\ that are supposedly not really detections at all. That this is not 
the case in the Morphs data is shown in Figure 8, where we have formed 
composite spectra by separating the sample into groups by their {\em
individual} measurements of \Hd. That \Hd\ grows stronger group
by group is, again, expected: this would happen even if we were only
measuring noise. More importantly, Figure 8 shows that, as \Hd\
increases in strength, the ratio of Ca H+H$\epsilon$ to Ca K also 
steadily increases, as do H$\gamma$ and the higher order Balmer lines. 
Even for the group $1.5 < \Hd\ < 3.0$, below the Morphs threshold 
for a k+a galaxy, several of the Balmer absorption lines are seen, at 
their expected strengths.  The obvious presence of many Balmer lines 
(clearly seen in Figure 8: H$\gamma$, H$\epsilon$, and H$\zeta$) 
in the composite spectra proves that the \Hd\ absorption lines 
identified in the Morphs spectra are genuine.

\subsection{Why the difference between CNOC1 and Morphs?}

The composite spectra presented here rule out one explanation
for the descrepancy between the Morphs and CNOC1 results, the
suggestion by B99 that the Morphs \Hd\ detections are spurious. 
Another of B99's suggestions relates to the statement in D99
that the Morphs selection of objects for spectroscopy was based
on morphology.  As it turned out, despite the Morphs's attempt to 
include a greater fraction of late-type galaxies (for those areas 
covered by the HST images), two features of the galaxy 
distribution --- the freedom to select
galaxies for a multislit mask is substantially limited by their spatial
distribution, and the probability that a target galaxy is either
foreground or background rises considerably with later Hubble 
type --- conspired to minimize the difference between the Morphs
selection and a pure magnitude-limited selection.  The biases
that resulted, discussed in P99, can be used to estimate how
much more likely was it that a galaxy with strong Balmer lines
(k+a, a+k, or e(a)) is included compared to a strictly 
magnitude limited sample.  The result is that, for the area 
covered by the HST exposures, there is a 23\% greater likelihood
that one of these types will be included.  When the remaining
area (for which no morphological information is available
was available and thus only magnitude--limited selection was 
used) is included drops this to well below 20\%.  This is probably 
significant, but much too small to explain the difference between 
the Morphs and CNOC1 studies in detecting galaxies with strong 
Balmer lines.     

Other than this selection bias, the only differences between the 
data samples that we have identified is that the Morphs reach about 
0.5 mag deeper ($R \sim 22.0$ at $z \sim 0.45$, the average redshift 
of the Morphs sample), and that the typical Morphs spectra have 
about twice the S/N of the typical CNOC1 spectra (by simple 
inspection) shown in Yee \et (1996).  Measurements of \Hd\ in 
particular are sensitive to quality of the data, because of the 
complicated and varying continuum shape around the feature, as
discussed earlier.

Apparently, then, both the Morphs and CNOC1 samples probe similar
depths to similar completeness. We have shown that our detection of
strong \Hd\ absorption in $z\sim 0.5$ cluster galaxies is genuine ---
why has it not been seen in the CNOC1 data?  An important first point
is that, although the total CNOC1 sample is very large, only 4 of the
clusters overlap the redshift range of the Morphs sample, $0.37 < z <
0.57$.  These 4 clusters contribute only about 300 cluster members to
the CNOC1 sample (see B99, Fig. 21),   Also, because of a S/N cut
applied by B99, not all the 300 are included in the search for k+a
galaxies.  The remaining sample over the relevant redshift range,
though unspecified, is clearly {\it smaller} than the 427 cluster 
members of the Morphs sample.

Second, B99 chose a higher threshold of 5 \AA\ for inclusion into the
k+a category, compared to the 3 \AA\ definition used by the Morphs,
which we have just shown by Figure 8 is supportable with the generally
higher S/N Morphs spectra.  This difference alone accounts
for a factor of two in the fractions of k+a galaxies, since $\sim$50\%
of the k+a/a+k galaxies in the Morphs cluster sample have 3 \AA\ $< \Hd
<$ 5 \AA.  Even with this more restrictive definition, B99 do find a
fraction of 4.4\% of k+a galaxies in their {\it entire} cluster sample,
equivalent to about 9\% with the Morphs definitions.  There is,
unfortunately, no breakdown specific to the Morphs redshift
range.\footnote{Instead, Figure 22 of B99 is used to show no 
growth in this fraction with increasing redshift, however, this plot 
was made after ``correcting'' the counts for supposedly large errors 
in the CNOC1 spectral measures.}

Third, there is a difference between the two studies of the 
method adopted to measure \EWHd.  We argue above that our method of
measuring \Hd\ line strengths by profile fitting is preferred over the
bandpass method used by the CNOC1 group, but over the stronger \EWHd\ $>$
5.0 \AA\ range used in CNOC1 the two techniques are in decent agreement,
as B99 also conclude. However, from the discussion in Section 3.2, we 
note that the bandpass technique significantly underestimates the
strength of \Hd\ when \Hd\ is weak, due to systematic changes in the 
continuum.  Because of this, the B99 distributions of \Hd\ strength are
significantly skewed to lower \EWHd, which leads to the 
derivation of a large scatter around zero \Hd\ strength. This effect 
is quite evident in Fig. 28 of B99, where the CNOC1 method is seen to substantially underestimate \Hd\ strength (applied to Morphs spectra 
and compared to Morphs data) for \EWHd\ $<$ 5.0 \AA.  As a result, 
the scatter in \Hd\ measurements is  {\it overestimated}, and this
becomes a key factor in B99's arguement that there is no statistically significant excess of k+a galaxies in the CNOC1 clusters.

Tran \et\ (2003) have reached the same conclusions regarding the CNOC1
measurements reported in B99, that there is a zero--point
shift and an overestimate of the errors in \Hd\ measurements.
For their own sample of high S/N spectra in three clusters $0.33 < z <
0.83$ (one common to B99) Tran \et\ find an intermediate fraction of
7--13\% of `E+A' (here k+a) galaxies, which they consider highly
significant and indicative of a major component of galaxy evolution in
clusters.

%
%
\hbox{~}
\centerline{\psfig{file=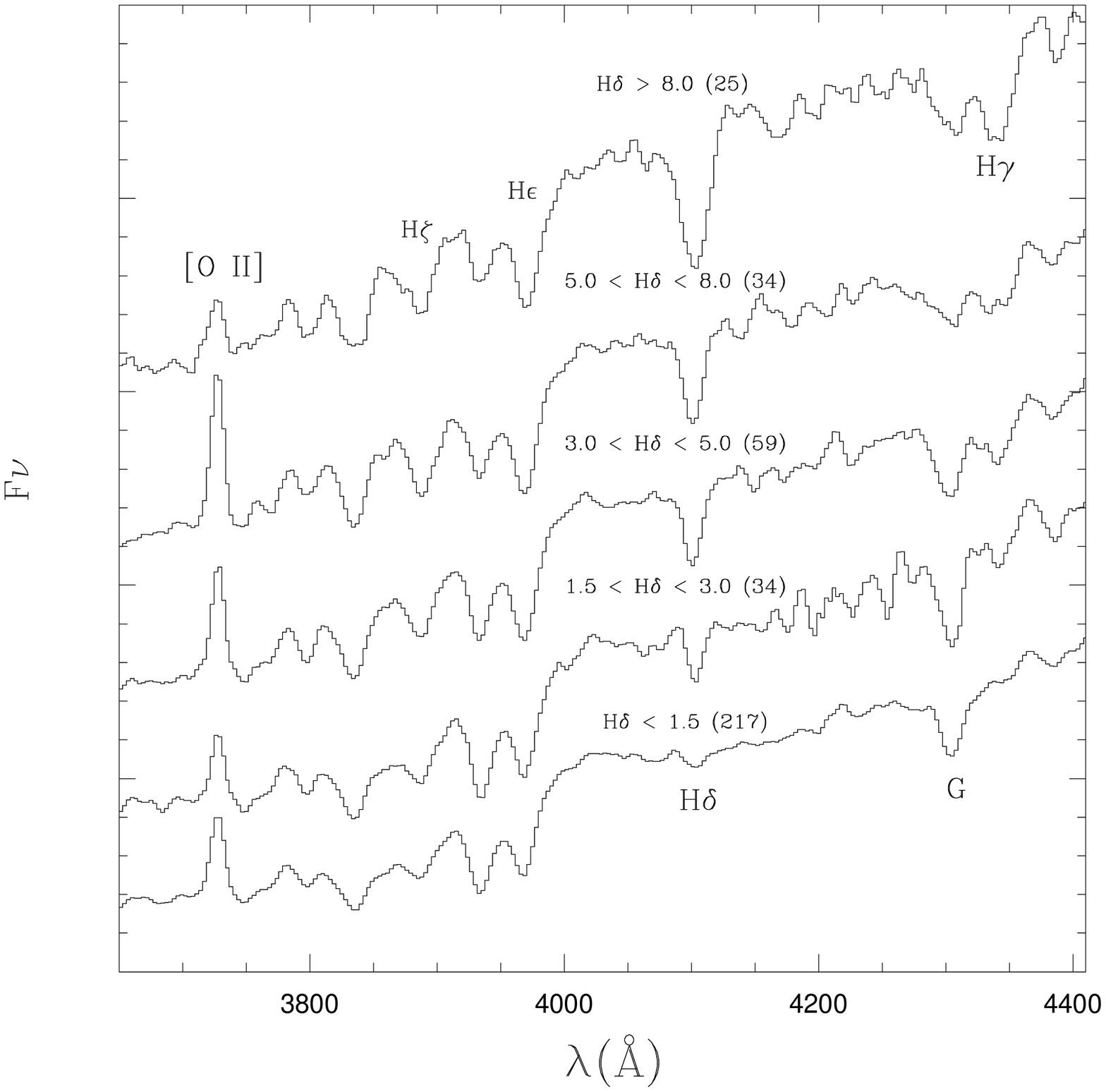,angle=0,width=5.5in}}
\noindent{\scriptsize
\addtolength{\baselineskip}{-3pt} 

\hspace*{0.3cm}Fig.~8.\ Composite spectra for Morphs sample galaxies
with \Hd\ absorption recorded in 5 ranges of strength.  The clear
trend of increasing \Hd\ strength and corresponding other members
of the Balmer series (note, in particular, H$\gamma$, H$\epsilon$,
and H$\zeta$ marked on top spectrum) shows the reality of the 
\Hd\ measurement, even in the presence of substantial noise in 
the continuum of the spectra of individual objects.

\vspace*{0.2cm}

\addtolength{\baselineskip}{3pt}
}

In summary, it is not clear that a discrepancy actually exists between
the Morphs finding of a significant fraction of k+a galaxies and the
results from CNOC1.  Differences in the quality of the data and its
treatment may explain much or even all of the disagreement.  If further 
work confirms a significantly different fraction of k+a galaxies in 
these two cluster samples, it could be evidence for the suggestion by
Ellingson \et\ (2001) that the difference in galaxy populations
reflects a difference in the two cluster samples.  Specifically, the
X--ray--selected clusters of CNOC1 include only rich, relaxed clusters,
while the Morphs clusters offer a wider diversity in properties, for
example, Abell 851, a cluster with strong subclustering that is
apparently undergoing rapid dynamical evolution.\footnote{For
comparison, the intermediate--redshift ($z > 0.35$) CNOC1 sample spans
a factor of 4.4 in X--ray luminosity compared to the Morphs range of
17.} If it is true that spectral evolution is correlated with cluster
characteristics, this would be a very helpful clue towards identifying
the mechanisms that drive the evolutionary phenomena which we observe.
 
Of course, a range in cluster properties could be an advantage
when investigating the evolution of cluster galaxies and
attempting to identify mechanisms of environmental influence.  It is
sensible to speculate that very luminous X--ray clusters at intermediate
redshift might have few k+a galaxies because they are dynamically the
most evolved clusters and further from their last major epoch of
cluster/group merging.  We note, however, that two very luminous X--ray
clusters, CL0016+1609 and 3C295 (see P99) are present in the Morphs 
sample, both of which contain significant number of k+a galaxies.
Because these are among the smallest samples in that study, we are
reluctant to conclude that the galaxy populations of these two
X--ray luminous clusters in the Morphs sample are as representative 
as those in the CNOC1 
clusters. Adding further confusion, MS1054 --- a very luminous
X--ray cluster --- appears to have a strong Balmer signature (see
Fig.~2), but at its high redshift of $z = 0.83$ this might indeed
reflect an earlier phase of the evolution of the most massive
clusters.  Until there are more and better data spanning a wide range
of X--ray luminosity for $z \sim 0.5$ clusters, we can certainly
not rule out the possibility that clusters with strong X--ray 
emission show the starburst phenomenon less prominently.

On the other hand, it worth noting that {\it all} of the Morphs
intermediate--redshift clusters, as well as the DGhiz sample and
MS1054, have a larger population of Balmer--strong galaxies than nearby
clusters, regardless of their masses/luminosities/morphologies
(Fig.~2). This suggests that cluster--to--cluster variations, at least
for clusters as rich as these and over this redshift range, are a 
second--order factor compared to evolutionary effects --- an important 
result on its own.  The Morphs sample of 10 clusters is still too 
small to demonstrate this conclusively or to make a statistically 
meaningful test for second--order effects.  Larger samples of 
clusters encompassing a wide range in X--ray luminosities/optical 
properties will be needed to assess the variation of the evolutionary 
trends with global cluster 
properties.  The RCS (Gladders \& Yee 2004), ACSCS, EDisCS (see www.mpa--garching.mpg.de/ediscs), and MACS (Ebeling, Edge, \&
Henry 2001) studies now underway will offer much--needed additional 
data.

\section{Summary}

We have shown the utility of composite spectra for consolidating the
evolution of cluster populations and significantly improving the
quality of diagnostic measurements.  Our choice of parameters
\EWHd\ and \EWOII\, which separate populations of very different
ages, makes it possible to dissect a cluster population by 
first adding the spectra of individual galaxies, thereby
overcoming a principal limiation of the data.

With this technique, we have demonstrated the reality
of the Morphs \Hd\ measurements, and have shown that the line--fitting
technique is the preferred method of measuring \Hd\ over a wide range of
\Hd\ strengths, and the high S/N of these composite spectra make it a
very robust technique. We have used these composite spectra to show
that the clusters in our sample separate cleanly between low and
higher redshift, with much stronger Balmer absorption for the latter.
Although the sample is small, this result suggests that cosmic
evolution has played a more important role in the evolution of cluster
galaxies than has the range in cluster properties.

We have also found that field galaxies at the same redshift also 
appear to have stronger Balmer lines than today's composite 
field galaxy populations.  By comparing to a present--epoch field
sample, the 2dF, we have shown that these higher redshift populations
cannot be made of any mixture of passive and continuously star forming
galaxies, and that starbursts must be occurring in a greater
fraction of these higher redshift galaxies. 

The high S/N of composite spectra enable the study of 
subtle features of the stellar populations of distant galaxies.
The technique of producing composite spectra of cluster populations
should also facilitate the construction of so--called Madau plots
that trace the evolution of star formation over time.  Although
these have been limited to field populations, it is clear that
an important step forward will be the construction of the
trend of star formation over epoch {\em and} environment (see,
for example, Postman, Lubin, \& Oke 2001, Bower \& Balogh 2004). 
As N--body simulations become better at predicting star formation 
along with the growth of dark--matter halos, data of this kind will 
put strong constraints on the models.

%
%
\clearpage
\begin{table*}
{\scriptsize
\begin{center}
\centerline{\sc Table 1}
\vspace{0.1cm}
\centerline{\sc Integrated Spectra }
\vspace{0.3cm}
\begin{tabular}{lcccccl}
\hline\hline
\noalign{\smallskip}
{ sample } & { total N } & { $<z>$ } & { \EWOII\ } & { \EWHd\ } \cr
\hline
\noalign{\smallskip}
 Abell  548    &   100   &  0.040  &       $-2.9$  &       1.2  \cr
 Abell   754  &   81   &  0.054  &     $>$ $-1.0$  &       0.6  \cr
 Abell  1631  &   71   &  0.053  &         $-1.8$  &       1.1  \cr
\hline
\noalign{\smallskip}
  Abell  1644  &   90   &  0.049  &       $-4.1$  &       1.5  \cr
  Abell  1736  &   96   &  0.041  &   $>$ $-1.0$  &       1.0  \cr
  Abell  1983  &   68   &  0.045  &       $-3.5$  &       1.5  \cr
\hline
\noalign{\smallskip}
  Abell  2151  &   41   &  0.037  &       $-4.6$  &       1.4  \cr
 DC0003-50     &   28   &  0.035  &       $-3.5$  &       0.9  \cr
 DC0428-53     &   79   &  0.041  &       $-1.6$  &       1.2  \cr
\hline
\noalign{\smallskip}
 DC0559-40     &  75    &  0.049  &       $-1.5$  &       1.1 \cr
 DC0608-33     &  13    &  0.035  &       $-7.9$  &       1.3 \cr
 DC2048-52     &  84    &  0.046  &       $-1.7$  &       1.0 \cr
\hline
\noalign{\smallskip}
 DSclusters    &  826   &  0.044  &       $-3.0$  &       1.1 \cr
 DSfield       &  130   &  0.044  &       $-8.4$  &       1.6 \cr
 DS\_Ellip     &  161   &  0.044  &   $>$ $-1.0$  &       0.9 \cr
\hline
\hline
\noalign{\smallskip}
 A370          &    40   &  0.374  &      $-3.9$  &       1.8  \cr
 CL3C295       &    18   &  0.459  &      $-3.7$  &       2.7  \cr
 CL0016        &    24   &  0.546  &      $-4.2$  &       2.4  \cr
\hline
\noalign{\smallskip}
 CL0024        &   97   &  0.393  &       $-8.3$  &       1.8  \cr
 CL0303        &   44   &  0.418  &       $-9.4$  &       1.8  \cr
 CL0939 (A851) &   84   &  0.406  &       $-5.4$  &       2.0  \cr
\hline
\noalign{\smallskip}
 CL1447        &    21   &  0.376  &      $-9.7$  &       2.3  \cr
 CL1601        &    51   &  0.539  &      $-2.3$  &       2.0  \cr
\hline
\noalign{\smallskip}
 Morphs        &   379   &  0.442  &      $-6.2$  &       2.0  \cr
 Morphs field  &    50   &  0.450  &     $-14.1$  &       2.7  \cr
 Morphs k      &   150   &  0.450  &  $>$ $-1.0$  &       1.2  \cr
\hline
\hline
\noalign{\smallskip}
 DGhiz         &    60   &  0.748  &      $-7.4$  &       2.0  \cr
 MS1054        &    72  &  0.803  &       $-4.1$  &       2.3  \cr
 DGhizfield    &    25   &  0.750  &     $-27.2$  &       3.7  \cr
 DGintzfield   &    23   &  0.450  &     $-24.8$  &       3.4  \cr
\noalign{\hrule}
\end{tabular}
\end{center}
}
\vspace*{-0.8cm}
\end{table*}

\end{document}